\documentclass[12pt]{iopart} 

\usepackage{graphicx}
\usepackage{subeqn}

\newcommand{\ket}[1]{|#1\rangle}

\begin{document}

\bibliographystyle{plain}

\title{Weak measurement and rapid state reduction in bipartite quantum systems}

\author{Charles Hill}
\ead{Charles.Hill@liverpool.ac.uk} \address{Department of Electrical
Engineering and Electronics, University of Liverpool, Brownlow Hill,
Liverpool, L69 3GJ, United Kingdom}

\author{Jason Ralph}
\ead{jfralph@liverpool.ac.uk} \address{Department of Electrical
Engineering and Electronics, University of Liverpool, Brownlow Hill,
Liverpool, L69 3GJ, United Kingdom}




\begin{abstract}
In this paper we consider feedback control algorithms for the rapid
purification of a bipartite state consisting of two qubits, when the
observer has access to only one of the qubits. We show 1) that the
algorithm that maximizes the average purification rate is not the
same as that that for a single qubit, and 2) that it is always
possible to construct an algorithm that generates a deterministic
rate of purification for {\em both} qubits.  We also reveal a key
difference between projective and continuous measurements with
regard to state-purification.
\end{abstract}

\maketitle



Although quantum measurements are often treated as instantaneous, in
practice measurement timescales can be significant when compared to
other relevant timescales. Recent experimental advances have meant
that it is now possible to observe continuous measurements of
individual quantum systems. The measurement of an observable is not
instantaneous, but takes place over a period of time \cite{Smi02,
Arm02, Ger04,Bus06}. This type of continuous measurement can be
modeled by considering a series of projective measurements on an
auxiliary system that is weakly coupled to the system of interest.
The auxiliary system, an environmental degree of freedom, is then
averaged out. This produces a continuous measurement record which
contains information about the evolution of the quantum system of
interest. The measurement record is then used to construct a best
estimate of the underlying evolution - which is referred to as an
`unraveling' of the master equation for the system \cite{Bel87,
Cav87, Car89, Kor99,Oxt05}.

When continuous weak measurement is applied, it is possible to
modify the evolution via \emph{Hamiltonian feedback}, where the
Hamiltonian applied to the system depends on the measurement record
\cite{Bel87, Kor99,Oxt05, Wis93, Hof98,Sto04,Wis94, Doh99, Com06}.
Hamiltonian feedback during measurement not only affects the final
state of the system,  but it can also also affect the rate of state
reduction. In a protocol described by Jacobs, the average rate of
state reduction (as measured by the purity of an initially mixed
state) for a single qubit is maximized \cite{Jac03}. This process is
known as rapid state reduction \cite{Com06}, or as rapid
purification \cite{Jac03}. Jacobs' protocol is deterministic, but
other protocols exist which are stochastic and minimize the average
time for a single qubit to reach a given purity \cite{Wis06}.

\begin{figure}
\begin{center}
\includegraphics[width=0.5\columnwidth]{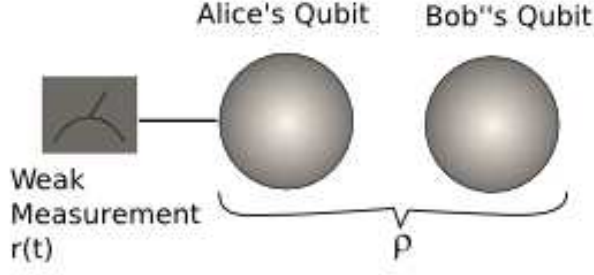}
\caption{Conceptual layout of the qubits. Alice and Bob both have a
single qubit. The bipartite system they share is described by the
density matrix $\rho$. Alice can make weak measurements on her
system, but not Bob's.} \label{fig:layout}
\end{center}
\end{figure}

In this paper we consider the analogous situation of performing
rapid state reduction in a two qubit system (shown in Fig.
\ref{fig:layout}). There are two parties, identified as Alice and
Bob, who may be separated and are not required to communicate. One
observer (Alice) has access to one qubit, which she can measure and
manipulate using local Hamiltonian feedback. She does not have
access to the second qubit, which is controlled by Bob. This
corresponds to a physically realistic situation in which Bob's qubit
is either spatially separated from Alice, or - for architectural
reasons - it is not possible to measure Bob's qubit directly. Two
qubit systems are important because they form the basis for many
current applications in quantum information processing, and are the
simplest system which exhibits entanglement.

In this paper we highlight three aspects of rapid state reduction
for bipartite qubit states: (1) We highlight a key difference
between projective measurements and weak measurement. (2) We show
that the measurement which provides the maximum rate of state
reduction for the unprobed qubit in a two qubit system is \emph{not}
necessarily the same as that for either of the known optimal single
qubit protocols \cite{Jac03, Wis06}. (3) We show that it is always
possible to purify both qubits deterministically at the same time.


The bipartite system which Alice and Bob share is described by the
density matrix $\rho$. Without loss of generality, Alice's system
undergoes a constant weak measurement, giving a measurement record,
$r(t)$.  It is natural to expand $\rho$ in the two-qubit Pauli
basis,
\begin{equation}
\rho = \frac{1}{4} \sum_{i,j = I,X,Y,Z} r_{ij} \sigma_{i} \otimes
\sigma_{j}, \label{eqn:rho}
\end{equation}
where $r_{ij}$ are real (since $\rho$ is Hermitian) and lie between
$-1$ and $1$, $\sigma_j$ are each of the three Pauli matrices and
the identity. Each $r_{ij}$ can be found, $r_{ij} = \Tr{\rho
\sigma_i \otimes \sigma_j}$.

The stochastic master equation (SME) which governs the evolution of
the density operator $\rho$ in the presence of a weak measurement of
a Hermitian observable, $y$, is given by
\begin{equation}
d\rho = -k[y,[y,\rho]] dt + \sqrt{2k}(y \rho + \rho y - 2 \langle y
\rangle \rho) dW. \label{eqn:SME}
\end{equation}
Here  $k$ is the measurement strength. The first term in this
equation describes the familiar drift towards the measurement axis.
The second term in the equation is weighted by $dW$, a Wiener
increment with $dW^2 = dt$. This term describes the update of
knowledge of the density matrix conditioned on the measurement
record \cite{Doh99}.

Throughout this paper we will consider measurements on Alice's qubit
alone. A measurement of Alice's qubit along a given axis $\hat{n}$
means that $y$ is given by
\begin{equation}
y = \hat{n} \cdot \sigma I = n_x XI + n_y YI + n_z ZI.
\end{equation}
Here, $X$, $Y$ and $Z$ are the Pauli matrices, and $I$ is the
identity. The tensor product is implied. The measurement direction,
$\hat{n}$ may change at each timestep. This is equivalent to the
application of single qubit Hamiltonian feedback to Alice's qubit,
except that the measurement axis rotates, rather than Alice's Bloch
vector.

Consider the evolution according to the SME as Alice's qubit
undergoes weak measurement along the $\hat{n}$ axis. The SME (Eqn.
(\ref{eqn:SME})) can be expressed in terms of the real Pauli
coefficients of the density matrix. If $\hat{n}, \hat{m}_1$ and
$\hat{m}_2$ are mutually orthogonal unit vectors, then the
corresponding stochastic master equation is given by
\begin{subequations}
\label{eq:dRn}
\begin{eqnarray}
dr_{mj} &=& -(4 k dt + r_{nI} \sqrt{8k}dW) r_{mj}, \\
dr_{nj} &=& (r_{Ij} - r_{nI}r_{nj}) \sqrt{8k} dW, \\
dr_{Ij} &=& (r_{nj} - r_{nI}r_{Ij}) \sqrt{8k} dW,
\end{eqnarray}
\end{subequations}
for $m = \{ \hat{m}_1, \hat{m}_2 \}$ and $j = \{X, Y, Z, I\}$. Here,
$r_{nj}$ is given by
\begin{equation}
r_{nj} = n_x r_{Xj} + n_y r_{Yj} + n_z r_{Zj},
\end{equation}
and similarly for $\hat{m}$. This is a system of 16 stochastic
differential equations.

In this paper, we will be particularly concerned with the evolution
of both Alice and Bob's reduced density matrices. Alice's reduced
density matrix is given by
\begin{equation}
\rho_A = \mathrm{Tr}_B \left( \rho \right) = \frac{I + r_{XI} X +
r_{YI} Y + r_{ZI} Z}{2}.
\end{equation}
Bob's reduced density matrix can be described by a similar equation
with the indices swapped.


The purity of a quantum state, $\rho$ can be quantified by the
purity, $P(\rho) =  \Tr{\rho^2}$. Purity has a minimum value of
$1/d$ where $d$ is the dimension of $\rho$, and a maximum value of
$1$. The purity of Alice's reduced density matrix is given by
\begin{equation}
P_A = \Tr{\rho_A^2} = \frac{1}{2}(1 + r_A^2), \label{eqn:PA}
\end{equation}
where $r_A$ is the Bloch vector of Alice's reduced density matrix. A
similar expression can be given for Bob's reduced density matrix.


Consider the rate of state reduction of Alice's qubit as she
measures on her own system. The change in average purity of Alice's
reduced density matrix given in Eqn. (\ref{eqn:PA}) under the
evolution of the SME, is given by
\begin{eqnarray}
dP_A &=& (1-r_{nI}^2)(1-r_A^2)4k dt + r_{nI}(1-r_A^2) \sqrt{8k} dW.
\label{eqn:dRn}
\end{eqnarray}
This expression does not depend on the state of Bob's qubit, and not
surprisingly, it is identical to the one qubit case. Hamiltonian
feedback can be used to implement either of the known single qubit
Hamiltonian protocols without modification \cite{Jac03, Wis06}.


Now we consider the opposite situation, when Alice would like to
find out the state of Bob's qubit. Bob's qubit is not being directly
measured. It is only through the correlations in the initial system,
$\rho$, that Alice can learn the state of Bob's system. For the most
effective purification, Alice and Bob's share an \emph{known}
initial entangled state.

The average change in purity (according to Alice) of Bob's qubit
when measured along the $\hat{n}$-axis is given by
\begin{eqnarray}
dP_B &=& [ (r_{nX} - r_{nI}r_{IX})^2 + (r_{nY} - r_{nI}r_{IY})^2 + (r_{nZ} - r_{nI}r_{IZ})^2 ] 4 k \mathrm{dt} +  \nonumber \\
& &[r_{IX} (r_{nX} - r_{nI}r_{IX}) + r_{IY} (r_{nY} - r_{nI}r_{IY})
+ r_{IZ} (r_{nZ} - r_{nI}r_{IZ}) ] \sqrt{8k} \mathrm{dW} ,
\end{eqnarray}
which can be written much more easily by identifying
\begin{subequations}
\begin{eqnarray}
r_{ni} &=& (r_{nX}, r_{nY}, r_{nZ}), \\
\Delta R_n &=& r_{nj} - r_{nI} r_{Ij}.
\end{eqnarray}
\end{subequations}
$\Delta R_n$ is a vector with three components, one for each of
$j=\{X,Y,Z\}$. Then the change in average purity of Bob's system
\begin{equation}
dP_B = \Delta R_n^2 4k dt + (r_{E} \cdot \Delta R_n) \sqrt{8k} dW.
\label{eqn:PE}
\end{equation}
We can write the nonlocal correlations as a matrix:
\begin{equation}
C_{ji} = r_{ij} - r_{iI} r_{Ij}, \label{eqn:C}
\end{equation}
for each $i,j = \{X,Y,Z\}$. To retrieve $\Delta R_n$ from $C$ we
multiply by the corresponding unit vector, $\hat{n}$. That is,
\begin{equation}
\Delta R_n = C \hat{n}.
\end{equation}
The magnitude square of $R_n$ gives the rate of state reduction of
Bob's qubit, as seen by Alice, when she makes a weak measurement
along the $\hat{n}$ direction.

We now consider how to maximize this rate, in a direct analogy to
the optimizing strategy used by Jacobs for a single qubit
\cite{Jac03}. In particular we find the largest average increase in
purity of Bob's reduced density matrix. Whilst there are some
circumstances in which this strategy will not give a globally
optimal solution, numerical simulations verify its use in this
application. The average increase in purity of Bob's reduced density
matrix is proportional to $|\Delta R_n|^2$. Expressing this in terms
of the matrix, $C$, we wish to find
\begin{equation}
dP_B^{(max)} = 4 k \max_{\hat{n}} |C \hat{n}|^2.
\end{equation}
The maximum value is given by the largest singular value,
$\sigma_{1}^2$, of $C$, therefore giving a maximum rate increase in
purity of $4 k \sigma_{1}^2$. The value of $\hat{n}$ which
corresponds to this maximum rate of state reduction is given by the
first right singular vector, $v_1$, of $C$.




Point (1) of this paper contrasts weak measurement and projective
measurement, using a specific example. The measurement which gives
the greatest increase in purity of Bob's qubit for a projective
measurement is not the same as the measurement which gives the
greatest rate of increase in purity of Bob's qubit for a weak
measurement.

\begin{figure}
\begin{center}
\includegraphics[width=0.8\columnwidth]{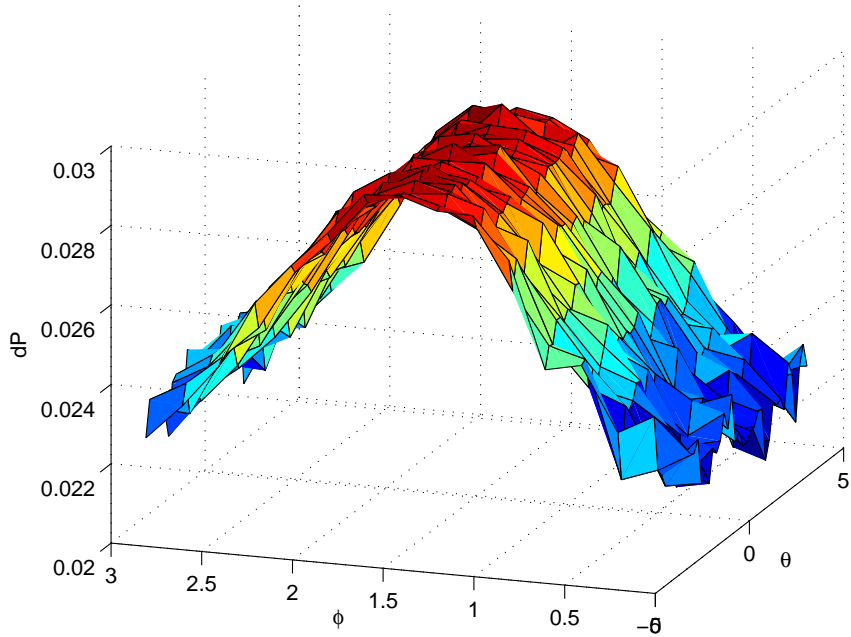}
\caption{Purification rate of $\rho_T$, with $\beta=0.5$, due to
weak measurement of Alice's qubit for every orientation of the
measurement axis after a dephasing of $\delta = 0.01$. $\phi$ is the
zenith angle, and $\theta$ is the azimuthal angle.}
\label{fig:0011Dephasing}
\end{center}
\end{figure}

Consider the state,
\begin{equation}
\ket{\psi} = \sqrt{\frac{1+\beta}{2}} \ket{00} +
\sqrt{\frac{1-\beta}{2}} \ket{11},
\end{equation}
where $-1 \le \beta \le 1$. $\ket{\psi}$ is the coherent
superposition, with both qubits aligned in the z-basis. If $\beta=0$
this is a maximally entangled state, and is entangled unless $\beta
= \pm 1$. If the system of interest undergoes a dephasing process,
as is common in many quantum systems, then the off diagonal terms of
the density matrix decay. If the amount of dephasing is
characterized by $\delta$ then the density matrix becomes
\begin{equation}
\rho_T = \frac{1}{4}\left(II + ZZ + \beta ZI + \beta IZ + \gamma
(XX-YY) \right),
\end{equation}
where $\gamma = \sqrt{1-\beta^2}-\delta$.

The maximum purity from a projective measurement is when Alice
measures along the z-axis. Any projective measurement by Alice along
the z-axis will project Bob's state into a pure state (either
$\ket{0}$ or $\ket{1}$), with purity, $P_B=1$. As Alice rotates the
measurement axis, the purity of Bob's state reduces. The matrix $C$
is
\begin{equation}
C = \left[
\begin{array}{ccc}
\gamma & 0 & 0 \\
0 & -\gamma & 0 \\
0 & 0 & 1-\beta^2
\end{array} \right].
\end{equation}
This matrix is diagonal. For small values of dephasing, $\delta$,
the maximum rate of state reduction occurs in the xy-plane, as is
shown in Fig. \ref{fig:0011Dephasing}. Therefore the projective
measurement which gives greatest purity is different from the weak
measurement which gives the greatest rate of increase in purity.



Point (2) of this paper contrasts weak measurement for bipartite
systems, and existing single qubit protocols. Naively, one might
expect that the best way for Alice to purify Bob's system is to
purify her own system fastest by applying known single qubit
protocols to her own qubit. We show, by giving a specific
counter-example, that this is not the case.

\begin{figure}
\begin{center}
\includegraphics[width=0.5\columnwidth]{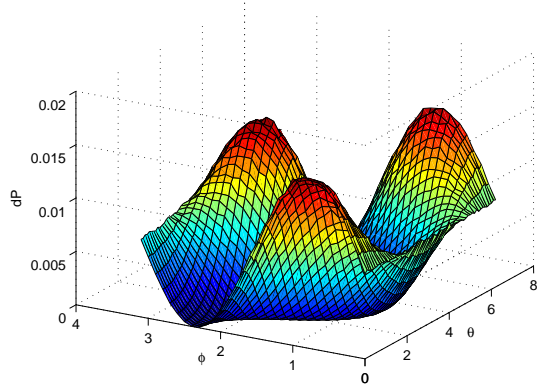}
\caption{Graph showing the rate of average state reduction of Bob's
qubit for all possible orientations of the measurement axis, and a
state given by $\rho_t$. $\phi$ is the zenith angle, and $\theta$ is
the azimuthal angle.} \label{fig:diffTo1Qubitb}
\end{center}
\end{figure}

Consider the state,
\begin{equation}
\rho_t =  \frac{1}{4} \left(II + \frac{1}{\sqrt{5}}(XI + XZ + ZZ)
\right).
\end{equation}
This state is perfectly aligned for Jacobs' one qubit feedback
protocol; Alice's reduced density matrix lies in the xy-plane
\cite{Jac03}. However, for the state $\rho_t$, the matrix $C$ is
given by
\begin{equation}
C = \left[
\begin{array}{ccc}
0 & 0 & \frac{1}{\sqrt{5}} \\
0 & 0 & 0 \\
0 & 0 & \frac{1}{\sqrt{5}}
\end{array} \right],
\end{equation}
which has a single non-zero singular value of $\sigma_1 =
1/\sqrt{5}$, and a corresponding vector of $n_1 = \frac{1}{\sqrt{2}}
(\hat{x} + \hat{z})$. We therefore expect the fastest rate of state
reduction for Bob's qubit occurs when Alice measures her own system
at $45^{\circ}$ to the z-axis and to the x-axis, as shown in Fig.
\ref{fig:diffTo1Qubitb}. The numerical values for this plot were
$k=0.1$, $dt=0.1$ and $N=10,000$ repetitions. This measurement does
not correspond to either the Jacobs' single qubit scheme
\cite{Jac03}, or the scheme proposed by Wiseman and Ralph
\cite{Wis06}. Purifying the unprobed qubit is dependant on the
correlations and the nature of those correlations. As we show here,
it is purified fastest by choosing a measurement axis to make use of
those correlations, and not by simply using a one qubit protocol.


Point (3) of this paper is that it is possible to make bipartite
state reduction deterministic. Increasing the purity of systems
deterministically is a desirable property. From a theoretical
standpoint it makes the equations easier, but the main advantages
are practical. If the evolution of purity is deterministic, then
each qubit is guaranteed to reach a set target purity in a given
time. When weak measurement is being used for state preparation in a
multiple qubit system, they will all reach the target purity
together.

The condition required for the purity of Alice's reduced density
matrix to evolve deterministically is that
\begin{equation}
r_A \cdot \hat{n} = 0. \label{eqn:aDet2}
\end{equation}
That is, the measurement should be taken in the plane orthogonal to
the direction described by Alice's reduced density matrix.


The condition for Bob's system to be deterministic is clear from
Eqn. (\ref{eqn:PE}). It is given by
\begin{eqnarray}
r_B' \cdot \hat{n} &=& 0, \label{eqn:determE}
\end{eqnarray}
where $r_B' = C^{T} r_{E}$. Therefore for the evolution of the
purity of Bob's reduced density matrix to be deterministic, the
measurement axis must be chosen in a plane orthogonal to $r_B'$.
This plane is spanned by $\hat{p}_1$ and $\hat{p}_2$.


Deterministic state reduction of Bob's qubit is desirable, but we
would still like to find the maximum rate of state reduction.
Although the measurement axis, $\hat{n}$ can be chosen anywhere in
the plane, not all orientations of $\hat{n}$ will purify Bob's
reduced density matrix at the same rate. The maximum deterministic
rate of state reduction of Bob's qubit is found by taking the
singular value decomposition of $CP_p$, where $P_p$ is given by
\begin{equation}
P_p = p_1 p_1^{T} + p_2 p_2^{T}.
\end{equation}

The fastest rate of average deterministic purification of Bob's
reduced density matrix is given by $4 k \sigma_{1}^2$ where
$\sigma_1$ is the largest singular value of the product $CP_p$, and
the axis of measurement is given by $\hat{n}_{det} = v_1$, the first
right singular vector of $C P_p$.


It is possible to choose a weak measurement which Alice can make on
her qubit so that the purity of both her reduced density matrix, and
also Bob's reduced density matrix both evolve deterministically. If
the measurement axis is chosen to be
\begin{equation}
n_{sim} \propto r_A \times r_B',
\end{equation}
then the evolution of the purity of both Alice and Bob's reduced
density matrices is deterministic.



In this paper we investigated the effect of weak measurements on a
bipartite system consisting of two qubits - Alice's qubit and Bob's
qubit. We allowed measurement on Alice's qubit alone. We gave
expressions for the rate of state reduction of both systems, based
on the measurement record. We have shown how to maximize the average
rate of state reduction of either system, and how to achieve
deterministic evolution of the purity. We have demonstrated that
weak measurement of two qubits can be very different from both
projective measurements and the weak measurement of a single qubit.
Many interesting effects occur in bipartite quantum systems under
measurement.


\ack

The authors would like to thank Kurt Jacobs and H. M. Wiseman for
his helpful comments and acknowledge that this work was supported by
UK EPSRC grant number EP/C012674/1.


\section*{References}
\bibliography{bibliographyLiverpool}

\begin{thebibliography}{28}
\expandafter\ifx\csname natexlab\endcsname\relax\def\natexlab#1{#1}\fi
\expandafter\ifx\csname bibnamefont\endcsname\relax
  \def\bibnamefont#1{#1}\fi
\expandafter\ifx\csname bibfnamefont\endcsname\relax
  \def\bibfnamefont#1{#1}\fi
\expandafter\ifx\csname citenamefont\endcsname\relax
  \def\citenamefont#1{#1}\fi
\expandafter\ifx\csname url\endcsname\relax
  \def\url#1{\texttt{#1}}\fi
\expandafter\ifx\csname urlprefix\endcsname\relax\def\urlprefix{URL }\fi
\providecommand{\bibinfo}[2]{#2}
\providecommand{\eprint}[2][]{\url{#2}}

\bibitem[1]{Smi02}
\bibinfo{author}{\bibfnamefont{W.~P.} \bibnamefont{Smith}},
  \bibinfo{author}{\bibfnamefont{J.~E.} \bibnamefont{Reiner}},
  \bibinfo{author}{\bibfnamefont{L.~A.} \bibnamefont{Orozco}},
  \bibinfo{author}{\bibfnamefont{S.}~\bibnamefont{Kuhr}}, \bibnamefont{and}
  \bibinfo{author}{\bibfnamefont{H.~M.} \bibnamefont{Wiseman}},
  \bibinfo{journal}{Phys. Rev. Lett.} \textbf{\bibinfo{volume}{89}},
  \bibinfo{pages}{133601} (\bibinfo{year}{2002}).

\bibitem[2]{Arm02}
\bibinfo{author}{\bibfnamefont{M.~A.} \bibnamefont{Armen}},
  \bibinfo{author}{\bibfnamefont{J.~K.} \bibnamefont{Au}},
  \bibinfo{author}{\bibfnamefont{J.~K.} \bibnamefont{Stockton}},
  \bibinfo{author}{\bibfnamefont{A.~C.} \bibnamefont{Doherty}},
  \bibnamefont{and} \bibinfo{author}{\bibfnamefont{H.}~\bibnamefont{Mabuchi}},
  \bibinfo{journal}{Phys. Rev. Lett.} \textbf{\bibinfo{volume}{89}},
  \bibinfo{pages}{133602} (\bibinfo{year}{2002}).

\bibitem[3]{Ger04}
\bibinfo{author}{\bibfnamefont{J.}~\bibnamefont{Geremia}},
  \bibinfo{author}{\bibfnamefont{J.~K.} \bibnamefont{Stockton}},
  \bibnamefont{and} \bibinfo{author}{\bibfnamefont{H.}~\bibnamefont{Mabuchi}},
  \bibinfo{journal}{Science} \textbf{\bibinfo{volume}{304}},
  \bibinfo{pages}{270} (\bibinfo{year}{2004}).

\bibitem[4]{Bus06}
\bibinfo{author}{\bibfnamefont{P.}~\bibnamefont{Bushev}},
  \bibinfo{author}{\bibfnamefont{D.}~\bibnamefont{Rotter}},
  \bibinfo{author}{\bibfnamefont{A.}~\bibnamefont{Wilson}},
  \bibinfo{author}{\bibfnamefont{F.}~\bibnamefont{Dubin}},
  \bibinfo{author}{\bibfnamefont{C.}~\bibnamefont{Becher}},
  \bibinfo{author}{\bibfnamefont{J.}~\bibnamefont{Eschner}},
  \bibinfo{author}{\bibfnamefont{R.}~\bibnamefont{Blatt}},
  \bibinfo{author}{\bibfnamefont{V.}~\bibnamefont{Steixner}},
  \bibinfo{author}{\bibfnamefont{P.}~\bibnamefont{Rabl}}, \bibnamefont{and}
  \bibinfo{author}{\bibfnamefont{P.}~\bibnamefont{Zoller}},
  \bibinfo{journal}{Phys. Rev. Lett.} \textbf{\bibinfo{volume}{96}},
  \bibinfo{pages}{043003} (\bibinfo{year}{2006}).

\bibitem[5]{Cav87}
\bibinfo{author}{\bibfnamefont{C.~M.} \bibnamefont{Caves}} \bibnamefont{and}
  \bibinfo{author}{\bibfnamefont{G.~J.} \bibnamefont{Milburn}},
  \bibinfo{journal}{Phys. Rev. A} \textbf{\bibinfo{volume}{36}},
  \bibinfo{pages}{5543} (\bibinfo{year}{1987}).

\bibitem[6]{Car89}
\bibinfo{author}{\bibfnamefont{H.~J.} \bibnamefont{Carmichael}},
  \bibinfo{author}{\bibfnamefont{S.}~\bibnamefont{Singh}},
  \bibinfo{author}{\bibfnamefont{R.}~\bibnamefont{Vyas}}, \bibnamefont{and}
  \bibinfo{author}{\bibfnamefont{P.~R.} \bibnamefont{Rice}},
  \bibinfo{journal}{Phys. Rev. A} \textbf{\bibinfo{volume}{39}},
  \bibinfo{pages}{1200} (\bibinfo{year}{1989}).

\bibitem[7]{Bel87}
\bibinfo{author}{\bibfnamefont{V.~P.} \bibnamefont{Belavkin}},
  \emph{\bibinfo{title}{Information, complexity, and control in quantum
  physics}} (\bibinfo{publisher}{Springer, New-York}, \bibinfo{year}{1987}).
\bibinfo{author}{\bibfnamefont{V.~P.} \bibnamefont{Belavkin}},
  \bibinfo{journal}{Rep. Math. Phys.} \textbf{\bibinfo{volume}{45}},
  \bibinfo{pages}{353} (\bibinfo{year}{1999}).
\bibinfo{author}{\bibfnamefont{L.}~\bibnamefont{Bouten}},
  \bibinfo{author}{\bibfnamefont{S.}~\bibnamefont{Edwards}}, \bibnamefont{and}
  \bibinfo{author}{\bibfnamefont{V.~P.} \bibnamefont{Belavkin}},
  \bibinfo{journal}{At. Mol. Opt. Phys.} \textbf{\bibinfo{volume}{38}},
  \bibinfo{pages}{151} (\bibinfo{year}{2005}).

\bibitem[8]{Kor99}
\bibinfo{author}{\bibfnamefont{A.~N.} \bibnamefont{Korotkov}},
  \bibinfo{journal}{Phys. Rev. B} \textbf{\bibinfo{volume}{60}},
  \bibinfo{pages}{5737} (\bibinfo{year}{1999}).
\bibinfo{author}{\bibfnamefont{A.~N.} \bibnamefont{Korotkov}},
  \bibinfo{journal}{Phys. Rev. B} \textbf{\bibinfo{volume}{63}},
  \bibinfo{pages}{115403} (\bibinfo{year}{2001}).
\bibinfo{author}{\bibfnamefont{R.}~\bibnamefont{Ruskov}} \bibnamefont{and}
  \bibinfo{author}{\bibfnamefont{A.~N.} \bibnamefont{Korotkov}},
  \bibinfo{journal}{Phys. Rev. B} \textbf{\bibinfo{volume}{66}},
  \bibinfo{pages}{041401} (\bibinfo{year}{2002}). 
\bibinfo{author}{\bibfnamefont{A.~N.} \bibnamefont{Jordan}} \bibnamefont{and}
  \bibinfo{author}{\bibfnamefont{A.~N.} \bibnamefont{Korotkov}},
  \bibinfo{journal}{Phys. Rev. B} \textbf{\bibinfo{volume}{74}},
  \bibinfo{pages}{085307} (\bibinfo{year}{2006}).

\bibitem[9]{Oxt05}
\bibinfo{author}{\bibfnamefont{N.~P.} \bibnamefont{Oxtoby}},
  \bibinfo{author}{\bibfnamefont{P.}~\bibnamefont{Warszawski}},
  \bibinfo{author}{\bibfnamefont{H.~M.} \bibnamefont{Wiseman}},
  \bibinfo{author}{\bibfnamefont{H.-B.} \bibnamefont{Sun}}, \bibnamefont{and}
  \bibinfo{author}{\bibfnamefont{R.~E.~S.} \bibnamefont{Polkinghorne}},
  \bibinfo{journal}{Phys. Rev. B} \textbf{\bibinfo{volume}{71}},
  \bibinfo{pages}{165317} (\bibinfo{year}{2005}).

\bibitem[10]{Wis93}
\bibinfo{author}{\bibfnamefont{H.~M.} \bibnamefont{Wiseman}} \bibnamefont{and}
  \bibinfo{author}{\bibfnamefont{G.~J.} \bibnamefont{Milburn}},
  \bibinfo{journal}{Phys. Rev. Lett.} \textbf{\bibinfo{volume}{70}},
  \bibinfo{pages}{548} (\bibinfo{year}{1993}).


\bibitem[11]{Hof98}
\bibinfo{author}{\bibfnamefont{H.~F.} \bibnamefont{Hofmann}},
  \bibinfo{author}{\bibfnamefont{G.}~\bibnamefont{Mahler}}, \bibnamefont{and}
  \bibinfo{author}{\bibfnamefont{O.}~\bibnamefont{Hess}},
  \bibinfo{journal}{Phys. Rev. A} \textbf{\bibinfo{volume}{57}},
  \bibinfo{pages}{4877} (\bibinfo{year}{1998}).

\bibitem[12]{Sto04}
\bibinfo{author}{\bibfnamefont{J.~K.} \bibnamefont{Stockton}},
  \bibinfo{author}{\bibfnamefont{R.}~\bibnamefont{{van Handel}}},
  \bibnamefont{and} \bibinfo{author}{\bibfnamefont{H.}~\bibnamefont{Mabuchi}},
  \bibinfo{journal}{Phys. Rev. A} \textbf{\bibinfo{volume}{70}},
  \bibinfo{pages}{022106} (\bibinfo{year}{2004}).

\bibitem[13]{Wis94}
\bibinfo{author}{\bibfnamefont{H.~M.} \bibnamefont{Wiseman}} \bibnamefont{and}
  \bibinfo{author}{\bibfnamefont{G.~J.} \bibnamefont{Milburn}},
  \bibinfo{journal}{Phys. Rev. A} \textbf{\bibinfo{volume}{49}},
  \bibinfo{pages}{1350} (\bibinfo{year}{1994}).
\bibinfo{author}{\bibfnamefont{J.}~\bibnamefont{Wang}} \bibnamefont{and}
  \bibinfo{author}{\bibfnamefont{H.~M.} \bibnamefont{Wiseman}},
  \bibinfo{journal}{Phys. Rev. A} \textbf{\bibinfo{volume}{64}},
  \bibinfo{pages}{063810} (\bibinfo{year}{2001}).
\bibinfo{author}{\bibfnamefont{H.~M.} \bibnamefont{Wiseman}},
  \bibinfo{author}{\bibfnamefont{S.}~\bibnamefont{Mancini}}, \bibnamefont{and}
  \bibinfo{author}{\bibfnamefont{J.}~\bibnamefont{Wang}},
  \bibinfo{journal}{Phys. Rev. A} \textbf{\bibinfo{volume}{66}},
  \bibinfo{pages}{013807} (\bibinfo{year}{2002}).

\bibitem[14]{Doh99}
\bibinfo{author}{\bibfnamefont{A.~C.} \bibnamefont{Doherty}} \bibnamefont{and}
  \bibinfo{author}{\bibfnamefont{K.}~\bibnamefont{Jacobs}},
  \bibinfo{journal}{Phys. Rev. A} \textbf{\bibinfo{volume}{60}},
  \bibinfo{pages}{2700} (\bibinfo{year}{1999}).
\bibinfo{author}{\bibfnamefont{A.~C.} \bibnamefont{Doherty}},
  \bibinfo{author}{\bibfnamefont{S.}~\bibnamefont{Habib}},
  \bibinfo{author}{\bibfnamefont{K.}~\bibnamefont{Jacobs}},
  \bibinfo{author}{\bibfnamefont{H.}~\bibnamefont{Mabuchi}}, \bibnamefont{and}
  \bibinfo{author}{\bibfnamefont{S.~M.} \bibnamefont{Tan}},
  \bibinfo{journal}{Phys. Rev. A} \textbf{\bibinfo{volume}{62}},
  \bibinfo{pages}{012105} (\bibinfo{year}{2000}).
\bibinfo{author}{\bibfnamefont{A.~C.} \bibnamefont{Doherty}},
  \bibinfo{author}{\bibfnamefont{K.}~\bibnamefont{Jacobs}}, \bibnamefont{and}
  \bibinfo{author}{\bibfnamefont{G.}~\bibnamefont{Jungman}},
  \bibinfo{journal}{Phys. Rev. A} \textbf{\bibinfo{volume}{63}},
  \bibinfo{pages}{062306} (\bibinfo{year}{2001}).

\bibitem[15]{Com06}
\bibinfo{author}{\bibfnamefont{J.}~\bibnamefont{Combes}} \bibnamefont{and}
  \bibinfo{author}{\bibfnamefont{K.}~\bibnamefont{Jacobs}},
  \bibinfo{journal}{Phys. Rev. Lett.} \textbf{\bibinfo{volume}{96}},
  \bibinfo{pages}{010504} (\bibinfo{year}{2006}).

\bibitem[16]{Jac03}
\bibinfo{author}{\bibfnamefont{K.}~\bibnamefont{Jacobs}},
  \bibinfo{journal}{Phys. Rev. A} \textbf{\bibinfo{volume}{67}},
  \bibinfo{eid}{030301} (\bibinfo{year}{2003})
\bibinfo{journal}{Proc. of SPIE} \textbf{\bibinfo{volume}{5468}},
  \bibinfo{pages}{355} (\bibinfo{year}{2004}).

\bibitem[17]{Wis06}
\bibinfo{author}{\bibfnamefont{H.~M.} \bibnamefont{Wiseman}} \bibnamefont{and}
  \bibinfo{author}{\bibfnamefont{J.~F.} \bibnamefont{Ralph}},
  \bibinfo{journal}{New J. Phys.} \textbf{\bibinfo{volume}{8}},
  \bibinfo{pages}{90} (\bibinfo{year}{2006}).
\bibinfo{author}{\bibfnamefont{H.~M.} \bibnamefont{Wiseman}}
  (\bibinfo{year}{2006}).

\end{thebibliography}

\end{document}